\documentclass{llncs}
\usepackage{graphicx}
\usepackage{amsmath}
\usepackage{amssymb}
\usepackage{enumerate}
\usepackage{varioref}
\usepackage{charter,eulervm}

\title{\sc {The Black-and-White Coloring Problem on 
Permutation Graphs}}

\author{
 Ton~Kloks
}
\institute{
 Department of Computer Science\\
 National Tsing Hua University\\
 Taiwan
}

\begin{document}

\maketitle

\begin{abstract}
Given a graph $G$ and integers $b$ and $w$. 
The black-and-white coloring problem asks if there 
exist disjoint sets of vertices $B$ and $W$ with $|B|=b$ and $|W|=w$ 
such that no vertex in $B$ is adjacent to any vertex 
in $W$. In this paper we show that the problem is polynomial 
when restricted to permutation graphs. 
\end{abstract}

\section{Introduction}

\begin{definition}
Let $G=(V,E)$ be a graph and let $b$ and $w$ be two integers. 
A black-and-white coloring of $G$ colors $b$ vertices black 
and $w$ vertices white such that no black vertex is adjacent to 
any white vertex. 
\end{definition}

In other words, the black-and-white coloring problem 
asks for a complete bipartite subgraph $M$ in the 
complement $\Bar{G}$ of $G$ with $b$ and $w$ vertices in the 
two color classes of $M$. 

\bigskip 

The black-and-white coloring problem is NP-complete 
for graphs in general~\cite{kn:hansen}. 
That paper also shows that 
the problem can be solved for trees in $O(n^3)$ time. 
In a recent paper~\cite{kn:berend} the worst-case timebound 
for an algorithm on  
trees was improved to $O(n^2 \log^3 n)$ time~\cite{kn:berend}. 
The paper~\cite{kn:berend} mentions, among other things, 
a manuscript by Kobler, 
{\em et al.\/}, which shows that the problem can be solved in 
polynomial time for graphs of bounded treewidth. 

\bigskip 

In this paper we investigate the complexity of the problem 
for permutation graphs. 
An intersection model for permutation graphs is obtained as follows. 
Consider two horizontal lines $L_1$ and $L_2$, one above the other. 
Label $n$ distinct points on $L_1$ and on $L_2$ with labels 
$\{1,\dots,n\}$. For each $k \in \{1,\dots,n\}$ connect the 
point with label $k$ on $L_1$ with the point with label $k$ on 
$L_2$ by a straight line segment. 
This is called a permutation diagram. The corresponding permutation 
graph with vertices $\{1,\dots,n\}$ is the intersection graph 
of the line segments. 

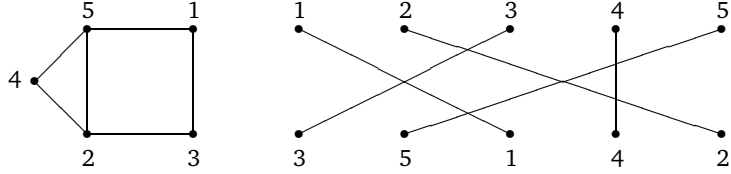
\begin{figure}
\setlength{\unitlength}{2pt}
\begin{center}
\begin{picture}(160,50)

\put(10,20){\circle*{1.5}}
\put(20,10){\circle*{1.5}}
\put(20,30){\circle*{1.5}}
\put(40,10){\circle*{1.5}}
\put(40,30){\circle*{1.5}}

\put(10,20){\line(1,-1){10}}
\put(10,20){\line(1,1){10}}
\put(20,10){\line(1,0){20}}
\put(20,10){\line(0,1){20}}
\put(40,10){\line(0,1){20}}
\put(20,30){\line(1,0){20}}

\put(5,19){$4$} \put(19,4){$2$}
\put(39,4){$3$} \put(19,32){$5$}
\put(39,32){$1$}

\put(60,10){\circle*{1.5}}
\put(80,10){\circle*{1.5}}
\put(100,10){\circle*{1.5}}
\put(120,10){\circle*{1.5}}
\put(140,10){\circle*{1.5}}

\put(60,30){\circle*{1.5}}
\put(80,30){\circle*{1.5}}
\put(100,30){\circle*{1.5}}
\put(120,30){\circle*{1.5}}
\put(140,30){\circle*{1.5}}

\put(60,30){\line(2,-1){40}}
\put(80,30){\line(3,-1){60}}
\put(100,30){\line(-2,-1){40}}
\put(120,30){\line(0,-1){20}}
\put(140,30){\line(-3,-1){60}}

\put(59,4){$3$}
\put(79,4){$5$}
\put(99,4){$1$}
\put(119,4){$4$}
\put(139,4){$2$}

\put(59,32){$1$}
\put(79,32){$2$}
\put(99,32){$3$}
\put(119,32){$4$}
\put(139,32){$5$}

\end{picture}
\end{center}
\caption{A permutation graph and its permutation diagram}
\label{figure permutation diagram}
\end{figure}

\medskip 

Permutation graphs can be recognized in linear time~\cite{kn:tedder}. 
A permutation diagram can be obtained in linear time.  
  
\section{Black-and-white colorings of permutation graphs}
\label{section permutations}

\begin{definition}
Consider a permutation diagram. A scanline is a linesegment 
that connects a point on $L_1$ with a point on $L_2$, such that 
the endpoints do not coincide with the endpoints of any line segment. 
\end{definition}

A black-and-white coloring with $b$ black vertices and 
$w$ white vertices is optimal if every uncolored vertex has a 
black and a white neighbor.  

\begin{lemma}
Assume there exists an  optimal black-and-white coloring of $G$ 
with $b$ black vertices and $w$ white vertices. 
There exists a collection of pairwise non-intersecting 
scanlines such that the vertices which are uncolored 
are exactly the line segments that cross one or more 
scanlines. 
\end{lemma}
\begin{proof}
Remove the line segments from the diagram of the vertices 
that are not colored. Each of the remaining components 
is colored black or white. Notice that the components 
form a consecutive sequence in the diagram. Place a scanline 
between any two consecutive components. The vertices that 
are not colored are precisely those that cross one of the 
scanlines. 
\qed\end{proof}

\bigskip 

\begin{theorem}
There exists a polynomial-time algorithm which checks 
if a permutation graph can be colored with $b$ black and 
$w$ white vertices. 
\end{theorem}
\begin{proof}
Consider a permutation diagram for a 
permutation graph $G=(V,E)$. A piece consists of a pair of 
non-intersecting scanlines. 

\medskip 

\noindent
Consider the subgraph of $G$ 
induced by the line segments with both endpoints between 
the two scanlines. Using dynamic programming, the algorithm 
checks if there is a black and white coloring of the piece 
with $b^{\prime}$ black and $w^{\prime}$ white vertices, 
for all values $b^{\prime}$ and $w^{\prime}$. We describe the 
procedure below.  

\medskip 

\noindent
The smallest pieces consist of two scanlines such that there 
is exactly one line segment between them. The subgraph induced 
by this piece has one vertex. There are two possible 
optimal colorings; 
either the vertex is black or it is white. 

\medskip 

\noindent
Consider an arbitrary piece, say that it is bordered by 
scanlines $s_1$ and $s_2$. 
Two possible colorings color 
the piece completely black or completely white. 
Cut the piece in two by a scanline $s$ which is between 
$s_1$ and $s_2$. Let $S$ be the set of line segments 
that cross $s$. The vertices of $s$ are uncolored. 
If there are colorings with $b_1$ and $w_1$ vertices 
colored black and white in the left piece and with 
$b_2$ and $w_2$ black and white vertices in the right 
piece, then the piece can be colored with $b_1+b_2$ 
black vertices and $w_1+w_2$ white vertices. 

\medskip 

\noindent
There are $O(n^4)$ different pieces, namely, there are 
$O(n^2)$ scanlines, and each piece is bordered by two of them. 
To process a piece, we try $O(n^2)$ scanlines that are between 
the two bordering scanlines. By table look-up, the algorithm 
checks if there is a black-and-white coloring of the 
piece with $b^{\prime}$ black vertices and $w^{\prime}$ 
white vertices. Each table consists of $O(n^2)$ entries. 
A table for the piece can be computed in $O(n^8)$ time. 
\qed\end{proof}
 
\begin{remark}
A similar algorithm can be obtained for the classes of 
circle graphs and $d$-trapezoid graphs~\cite{kn:bodlaender,kn:kratsch}. 
\end{remark}

\end{document}